# Jupyter notebooks as discovery mechanisms for open science:
# Citation practices in the astronomy community

Morgan F. Wofford, Bernadette M. Boscoe, Christine L. Borgman, Irene V. Pasquetto, and Milena S. Golshan

**Abstract**— Citing data and software is a means to give scholarly credit and to facilitate access to research objects. Citation principles encourage authors to provide full descriptions of objects, with stable links, in their papers. As Jupyter notebooks aggregate data, software, and other objects, they may facilitate or hinder citation, credit, and access to data and software. We report on a study of references to Jupyter notebooks in astronomy over a 5-year period (2014-2018). References increased rapidly, but fewer than half of the references led to Jupyter notebooks that could be located and opened. Jupyter notebooks appear better suited to supporting the research process than to providing access to research objects. We recommend that authors cite individual data and software objects, and that they stabilize any notebooks cited in publications. Publishers should increase the number of citations allowed in papers and employ descriptive metadata-rich citation styles that facilitate credit and discovery.

**Index Terms**—Astronomy, software, data citation, Jupyter notebooks, credit, attribution, discovery, open science

---

## 1 INTRODUCTION

IMPROVING access to data and software associated with publications is a central tenet of the open science movement. However, sharing data and software is a labor- and resource-intensive process of documenting, curating, and depositing research objects. Disincentives are many, but among the incentives is the opportunity to obtain credit for data and software as scientific contributions [1], [2]. In a technologically ideal world, all research objects – "semantically rich aggregations of resources that provide 'units of knowledge'" [3] – such as publications, code, datasets, and documentation, would be identified uniquely and linked to all related objects in a dense and searchable network.

Today's reality of scholarly communication is far from that elegant network of typed relationships among scientific products. Research objects proliferate in units of many types and sizes, with widely varying degrees of intellectual control. Publications are the most carefully documented, with metadata and provenance information in the form of bibliographic citations. Principles and practices for data citation have emerged over the last decade, followed by proposals for software citation principles.

Efforts to map bibliographic citation practices to data and software encounter the problems of fixity and granularity [4]. Fixity is a notion central to citation practice and to copyright law. To be citable, an object should exist in a fixed form that can be retrieved in the same form or state as it was when referenced. Granularity is the notion that cited objects exist in a stable and definable unit, usually with clear boundaries.

Scholarly publications are canonical examples of fixity and granularity, although often imperfect. Publications such as journal articles, books, and conference papers are distinct objects whose characteristics can be described with a basic set of metadata elements such as author, title, date, journal, publisher, and so on (setting aside the complexities of historical documents, grey literature, and bibliographic theory). While versions may proliferate, once an article or book is published, it usually stays published in a fixed form. Granularity of publications can break down when tables and figures within journal articles are assigned their own Digital Object Identifiers (DOIs), as is now the case. Similarly, references can be made to papers within conference proceedings, chapters within edited volumes, and whole proceedings and volumes.

Fixity and granularity problems are much more acute in data and software. They are less readily described by a core set of metadata elements than are publications,

---

- M. F. Wofford is with the UCLA Department of Information Studies, Los Angeles, CA 90095. E-mail: mfwofford@ucla.edu.
- B. M. Boscoe is with the UCLA Department of Information Studies, Los Angeles, CA 90095. E-mail: boscoe@ucla.edu.
- C. L. Borgman is with the UCLA Department of Information Studies, Los Angeles, CA 90095. E-mail: christine.borgman@ucla.edu.
- I. V. Pasquetto is with the Harvard Shorenstein Center of the Kennedy School of Government, Cambridge, MA 02138. E-mail: irene_pasquetto@hks.harvard.edu.
- M. S. Golshan is with the Los Angeles County Museum of Art, Los Angeles, CA 90036. E-mail: milenagolshan@ucla.edu.

although efforts to do so are proliferating. A dataset associated with a journal article or paper may be part of larger datasets and may exist in multiple versions. Similarly, software encompasses packages that change versions or releases frequently and that may include code, scripts, and algorithms associated with hardware, software, and data that are subject to local configurations. DOIs and other identifiers are being assigned to these research objects, but minting an identifier does not stabilize the object it describes. Fixity is usually achieved by depositing a research object in a public repository that curates and maintains it in stable form.

Digital "laboratory" notebooks may help to tame this morass by aggregating research objects associated with a project or publication, such as data, software, and code. Jupyter notebooks are a tool that has been adopted widely in scientific communities to improve reproducibility, reuse of data and code, and scientific transparency by aggregating research objects. The questions addressed in this paper concern how, and how well, Jupyter notebooks function as data or software citation mechanisms. By bringing together related objects, they could provide a means to discover, access, and give credit for the objects they contain. Conversely, they could "black box" those objects, making them more difficult to discover individually. A related question is how robust are Jupyter notebooks' mechanisms for discovery, attribution, credit, and access.

## 2 Background and Literature Review

As a basis for exploring how a new tool such as Jupyter notebooks might facilitate credit, attribution, and discovery of research objects, this section provides a brief overview of principles for citing publications, data, and software, and examines the Jupyter notebook as a research object.

### 2.1 Citation Principles and Practices

Bibliographic citations have a long history in scholarly communication. Strictly speaking, a citing paper makes a reference to cited paper; references are given and citations are received. The citation creates a directional relationship between the two publications. Bibliometrics is the study of these relationships for purposes such as the influence of authors, journals, disciplines, or countries; flows of ideas between communities or over time; or the growth and evolution of disciplines. Citations serve functional purposes such as establishing evidentiary sources, giving credit, and facilitating the discovery and retrieval of materials on which the citing publication is based [5].

To serve these functions, bibliographic references must contain adequate metadata to identify the cited item uniquely. References to published documents normally include elements such as author, title, journal, book title and publisher, date and place of publication, and volume, issue, and page numbers. The latter three data elements position the published object in the bibliographic universe of print publications. For objects in electronic form, descriptive metadata are augmented by a unique and persistent identifier, such as a Digital Object Identifier (DOI), to position it in the bibliographic universe of digital publications. While bibliographic citation practices are adequate for most cited items to be discovered and retrieved, citation formats are far from robust or standardized. Bibliographic citation styles number in the thousands, unique and persistent identifiers may be neither, and citing authors create noisy bibliographic data by failing to verify details such as middle initials, dates, page numbers, DOIs, and URLs.

Citations, thus, are much more than links between objects. To serve the scholarly communication functions of giving credit, and to enable the reader of the citing document to discover and retrieve the cited document, descriptive metadata are necessary. In the case of digital objects, a valid hyperlink between the objects is necessary but not sufficient. Hyperlinks, whether URLs, DOIs, or other pointers, lack descriptive information and are subject to "link rot." Thus, our research design makes careful distinctions between citations and hyperlinks.

### 2.2 Data Citation Principles

Data citation principles arose for purposes similar to those for bibliographic citation – the need for credit, attribution, and discovery. CODATA convened an international task group on Data Citation and Attribution that convened workshops and issued reports [6], [7]. Multiple stakeholders came together to synthesize the CODATA and similar efforts into the Joint Declaration of Data Citation Principles, which settled on eight core principles: Importance, Credit and Attribution, Evidence, Unique Identification, Access, Persistence, Specificity and Verifiability, and Interoperability and Flexibility [8], [9]. These eight principles provide the basis for citation styles that may vary by domain, publisher, type of data, and other factors. The importance principle calls for data used in publications to be referenced on a par with that of cited publications. The credit and attribution principle indicates that sources of data, including individuals and institutions, are to be identified with appropriate metadata, akin to author, title, date, and other elements of bibliographic citations. The unique identification principle is usually implemented with a DOI, Handle, or domain-specific identifier. Persistence implies that the dataset and references to it will remain accessible for some period of time, perhaps indefinitely. This principle also assumes that cited datasets are stable and fixed objects.

### 2.3 Software Citation Principles

Software citation principles have been slow to develop due to challenges of fixity and granularity. To address these challenges, the FORCE11 Software Citation Working Group selected six of the eight principles for data citation, slightly renamed [10]. These six principles are Importance, Credit and Attribution, Unique Identification, Persistence, Accessibility, and Specificity. Together, they promote best practices that address the unique characteristics of software such as using a DOI associated with a specific software release to implement the unique identification principle [4]. Software packages, including code, scripts, and algorithms, are executable entities, making them difficult to describe as fixed digital objects [11], [12]. From a computational science perspective, each software layer depends upon the robustness of the all the layers below it. In Hinsen's model [13] (Figure 1), the top two layers are domain-specific, the third is a general scientific layer, and the fourth is a non-scientific software layer. Below this stack lie the hardware and the operating systems on which computations and basic services are based. Changes or instability in any layer undermines the stability of the layers above, in what Hinsen calls "software collapse" [13]. Therefore, references to software at any of these layers can only be as unique, persistent, and accessible as the stability of software in the layers below.

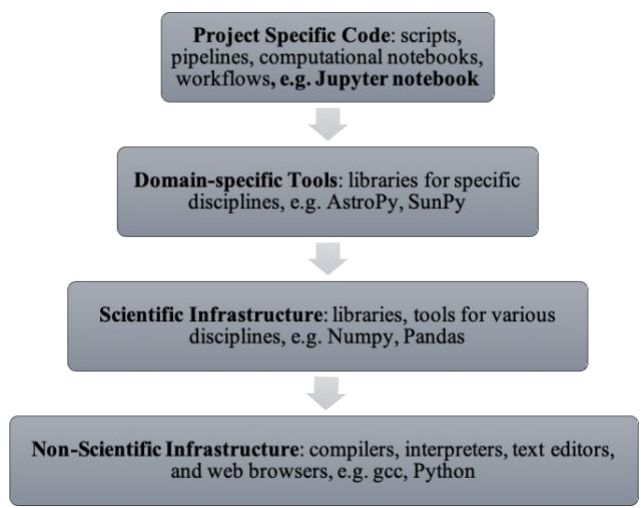

Fig. 1. The Computational Science Software Stack. Adapted from [12]

## 2.4 Jupyter Notebooks

Jupyter notebooks are open source web applications that act as virtual lab notebooks to analyze datasets, create visualizations, display computations, and more. The Jupyter notebook environment consists of an underlying JSON notebook, the server-client application, and a notebook kernel that executes the code in the JSON notebook. This structural simplicity allows a researcher considerable flexibility to employ Jupyter notebooks. Known as iPython notebooks until 2014, Jupyter notebooks were developed to accomplish faster, "on the fly" computing for data exploration, saving considerable time and labor over traditional approaches to compiling code [14]. However, as their popularity grew, researchers began to use Jupyter notebooks to support all the computational phases of the research lifecycle including data exploration, publication, and education [15].

The Jupyter notebook allows users to aggregate research objects such as software, data, methods, and contextual information to create an integrated computational workflow that is greater than the sum of its parts. A growing number of researchers reference Jupyter notebooks in their publications. When a Jupyter notebook is made available to others, users can access datasets via APIs, run and rerun code snippets, show results, and view documentation of processes and procedures. Jupyter notebooks commonly reside in open repositories such as GitHub.

Jupyter notebooks are research objects that aggregate other research objects, and thus are not easily categorized as data, software, or publications for the purposes of citation. We place Jupyter notebooks at the top of Hinsen's computational software stack (Figure 1) [13] because it relies on the lower layers for stability. A Jupyter notebook can function as a single entry point to all research objects associated with a research project or a paper. Ideally, if any one of these objects are discovered, a path can be followed via the notebook to related objects via hyperlinks, or code such as an API call. However, if individual research objects are described only within a Jupyter notebook, they may not be discoverable on their own, becoming "black boxed" within the notebook.

## 2.4 Case Study in Astronomy

Astronomy is a data-intensive domain with a longer history of open access to data and software than most sciences. The community invests in a collaborative international infrastructure of shared instruments, data standards, digital archives, metadata and discovery services [16]. However, even with this established infrastructure, data and software citation practices have been inconsistent and subject to link rot [17]. Starting in 2015, the American Astronomical Society (AAS), which is the major publisher for the field, and the Mikulski Archive for Space Telescopes (MAST), one of the primary data archives in astronomy, began a pilot project to assess the feasibility of implementing data citation across their journals and archives [18]. The pilot establishes mechanisms for depositing datasets in MAST at the time of submitting manuscripts to AAS journals, and obtaining a DOI for the dataset as part of the submission processes. Fixity is achieved by depositing the dataset in MAST. Granularity is addressed by scoping the dataset associated with the article, rather than referencing an entire sky survey, for example. The pilot study also emphasizes the importance of binding the paper and the dataset, as the former is necessary to interpret the latter.

Astronomers also are heavy users of Python and Jupyter notebooks. Among the projects employing Jupyter notebooks are the Large Synoptic Survey Telescope, Hubble Space Telescope, Chandra X-Ray Observatory, Sloan Digital Sky Survey, and the Square Kilometre Array [15], [19].

## 3 METHODS

Because Jupyter notebooks aggregate research objects, they may facilitate or obscure access to the objects they contain. Our method of determining the degree to which Jupyter notebooks facilitate data and software citation is to inspect all records containing the term "Jupyter" in astronomy papers since their inception in 2014. We chose the domain of astronomy both because they are heavy adopters of Jupyter notebooks and because the literature of the field is comprehensively indexed in one source, the Astrophysics Data System (ADS) [20].

In March 2019, we queried ADS via the "Bumblebee" interface using full text search for "Jupyter" (full:"Jupyter"), which is a distinctive term. As of July 2019, ADS contains more than 14.6 million records, indexed from publishers, arXiv, and other sources that are relevant to astronomy and astrophysics, including selected areas of physics, chemistry, computer science, mathematics, and geophysics. ADS records vary in scope and content, ranging from an independent abstract of a presentation to a bibliographic reference with metadata such as title, authors, abstracts (if available), links to the paper or presentation (if it exists), and other information that varies by the research object's origin.

We applied several filters to the retrieved set, as explained in the findings. First, we sorted records by whether they were solely abstracts or whether they contained descriptions of papers, preprints, articles, or other publications. Second, we retrieved any papers linked from the record. If both the published version and preprint version of a paper were available in a record, we used the published version for subsequent stages of analysis. Third, we sorted the remaining records by whether the paper contained a hyperlink to one or more Jupyter notebooks or whether they merely mentioned the term "Jupyter" in the paper (e.g., "we used a Jupyter notebook for our data analysis"). Fourth, we attempted to retrieve Jupyter notebooks from records that contain links, following those links to repositories, project websites, and NBviewers. When multiple notebooks or repositories were cited in one paper, we coded in combination (i.e. NBviewer + GitHub). We classified the records by whether the Jupyter notebooks could be obtained. To be classified as findable and accessible, the paper to which the record referred had to satisfy three conditions: 1) a link to the Jupyter notebook or repository was valid, meaning the research object at the destination could be obtained; 2) the Jupyter notebook was available at the destination without extra permissions; and 3) the notebook was available from the link and could be opened.

Our fifth filter, applied to the subset of records/papers that led to findable and accessible Jupyter notebooks, was to classify them by whether the reference contained only a hyperlink or whether it was a reference that also included metadata as recommended in the data and software citation principles. Those references might include metadata such as names of creators, dataset descriptions, versions of datasets or software, names of software packages, code dependencies, and other descriptive information. While Jupyter notebooks predate the formal publication of data and software principles, the research question of this article addresses how well these notebooks facilitate access to data and software. Sixth and last, we made general note of the authors' expressed reasons for referencing Jupyter notebooks. The reasons were too complex to be classified reliably, as we determined in an earlier pilot study [21].

## 4 RESULTS

We report results in five categories, based on the six filtering and coding steps outlined in the methods section. The first category reports the filtering from all occurrences of "Jupyter" to those records containing a link to a Jupyter notebook. The second category is to filter records whose links meet the three criteria for findable and accessible notebooks; the remaining analyses are based on these papers and notebooks. The third category identifies locations where those notebooks were found. In the fourth category, we distinguish between records that contain only links and those that provide a citation to data or software. Lastly, we document reasons that authors give for using Jupyter notebooks in astronomy and astrophysics projects.

### 4.1 How are Jupyter notebooks referenced in astronomy publications?

Figure 2 is a graph of the distribution of ADS records retrieved for the years 2014 through 2018. A total of 897 ADS records in this time frame contained the term "Jupyter." The ADS records referencing Jupyter notebooks first appear in 2014, when the iPython notebook became the Jupyter notebook. Only 3 records appeared in 2014, steadily increasing to 511 records in 2018. The graph shows the growth in usage of Jupyter notebooks in the astronomical community, as indicated by the occurrence of the term in ADS records, papers, and publications.

The black solid line (top) in Figure 2 represents all the ADS records with Jupyter notebook references (n=897). The dark gray dashed line (middle) represents ADS records with Jupyter notebook references that link to papers (n=755). The remaining 142 ADS records (897-755) were simply abstracts of talks or papers; they did not contain links or references to any external materials. The light gray dotted line (bottom) represents ADS records with papers containing links to Jupyter notebooks and to the repositories in which the notebooks reside (n=431). The remaining 342 records (755-431) linked to papers or publications, but the referenced documents did not contain hyperlinks or citations to Jupyter notebooks.

In sum, fewer than half (431 of 897 =48%) of the records identified with a full text search for "Jupyter" contained a link to one or more notebooks. Whereas the number of occurrences of the term increased steadily, the proportion of referenced notebooks remained about half each year.

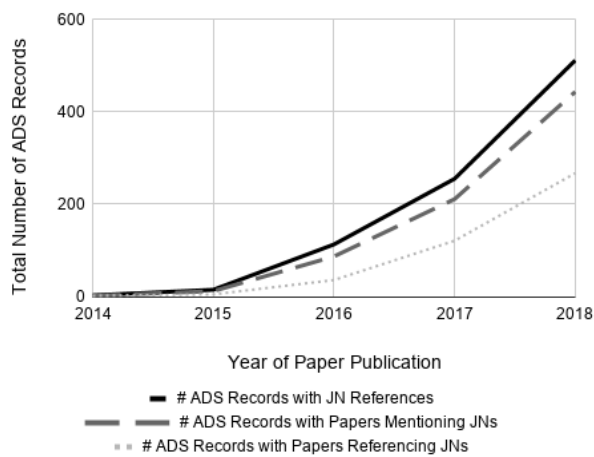

Fig. 2. ADS Records and Papers with Jupyter Notebooks (JN) References and Links 2014-2018. Coded by paper and Jupyter notebook link availability.

### 4.2 How stable are the referenced Jupyter notebooks?

For the 431 papers containing links to a Jupyter notebook, we tested for the three conditions of being findable and accessible:
1. A link to the Jupyter notebook or repository was valid, meaning the research object at the destination could be obtained.
2. The Jupyter notebook was available at the destination without extra permissions.
3. The notebook was available from the link and could be opened.

In total, 392 of the 431 led to findable and accessible notebooks. The remaining 39 (9%) had dead links due to the endemic "link rot" problem [17] or required additional permissions for access. While link rot would suggest a steady decline in accessibility by year, the pattern was not consistent. One of the two links in 2014 was valid, all 5 of the 2015 links were valid, whereas 8 of 36 links in 2016 were invalid, 9 of 121 in 2017 were invalid, and 21 of 267 links in 2018 were invalid.

### 4.3 Where are referenced Jupyter notebooks located or stored?

In following the links in the 392 papers that led to findable and accessible Jupyter notebooks, we categorized where the Jupyter notebooks were stored or, in the case of NBviewer, displayed. Of the 392, 264 (67%) papers linked solely to a GitHub repository containing the notebooks and associated files including other research objects like data and code. In 20 (5%) of the papers, a DOI issued by Zenodo linked to Jupyter notebooks and to a GitHub repository. Another 18 (5%) papers provided direct downloads of Jupyter notebooks in the form of a zip file or tar ball, while 16 (4%) linked only to a project website. NBviewer, which compiles and displays the notebook as static webpages, was the link destination for 15 (4%) of the 392 papers and publications. The remaining 59 (15%) of the links led to other destinations such as Figshare, Bitbucket, MyBinder, other repositories, and combinations thereof. GitHub was overwhelmingly the preferred destination and storage option for Jupyter notebooks.

In comparing the link destinations for the 39 papers for which Jupyter notebooks could not be found, project websites were 15 of these, another 7 were links to GitHub alone, 4 were to direct downloads, one each to Zenodo and NBViewer, and 11 to the other destinations category. Thus, as a proportion of links, project websites provided the least fixity, followed by direct downloads and assorted other repositories. While we were able to retrieve notebooks from GitHub and these other sites, we could not test whether the notebooks were unchanged from the time that the reference was made.

### 4.4 How do Jupyter notebooks receive links and citations?

In our sixth filtering step, we coded each paper based on the type of Jupyter notebook references provided, distinguishing citations, that is, references that contain links and associated metadata (e.g., creators, dataset descriptions, versions, software releases, packages, dependencies) and records that contain only links (e.g., URLs, DOIs, Direct Download), lacking associated metadata. Figure 3 shows this distribution. Of the 392 papers leading to findable and accessible Jupyter Notebooks, 105 include sufficient metadata to be considered a citation, whereas 287 contain only links. The grey (lighter) portion of each bar is the number of papers that provide only links to reference Jupyter notebooks or repositories. The black (darker) portion is the number papers that use citations to reference Jupyter notebooks or repositories. While the ratio of papers providing citations to Jupyter notebooks increases over time, starting at 0% in 2014 and peaking at roughly 29% in 2018, the average is 27% of findable and accessible notebooks receiving citations.

Of the 39 links for which notebooks were not findable and accessible, 33 were links alone and 6 were full citations, affirming the greater fixity of citations.

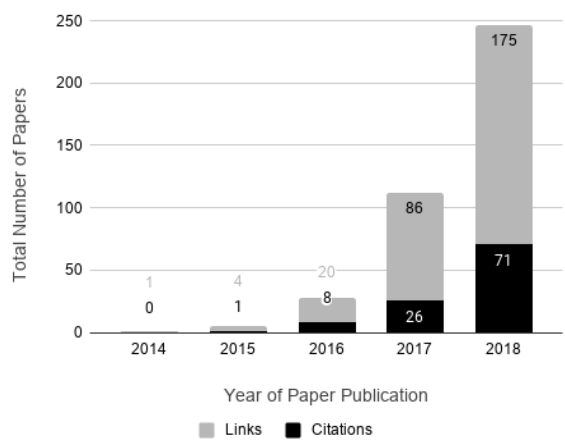

Fig. 3. Citation Practices for Jupyter Notebooks from ADS Papers 2014-2018. Grey bars represent links to Jupyter notebooks, while black bars represent citations with links and additional metadata. N= 329 records that contain papers with findable and accessible links.

Locations of references to Jupyter notebooks varies widely, occurring in the text, supplemental material, footnotes, or reference lists. Links often occur only as mentions in the text of an article:

> "We additionally provide a series of python libraries and jupyter notebooks with the computer code we used for the analysis on github: https://github.com/kgullikson88/BinaryInference." (Source: https://doi.org/10.3847/0004-6256/152/2/40)

In other papers, links are buried in supplementary material; this link initiates a direct download of a Jupyter notebook:

> "SUPPLEMENTAL MATERIAL: example-homodyne-Jaynes-Cummings-system.ipynb" (Source: https://doi.org/10.1103/PhysRevApplied.7.044002)

Citations also vary considerably in format and location. Principles for data and software citation encourage inclusion in reference lists, such as this one, which links to NBviewer along with creator, data, and other metadata.

> "[9] P. Norvig (2014) xkcd 1313: Regex Golf. Retrieved 2017-11-10. [Online]. Available: http://nbviewer.jupyter.org/url/norvig.com/ipython/" (Source: http://arxiv.org/abs/1712.07671)

In other cases, citations were found in tabular form in an appendix to a paper, as in this example:

TABLE 1
CITATIONS IN AN APPENDIX

| | |
|---|---|
| Current code version | v0.2.0 |
| Permanent link to code/repository used for this code version | https://github.com/ElsevierSoftwareX/SOFTX-D-17-00073 |
| Legal Code License | UnLicense (Public Domain) |
| Code versioning system used | git |
| Software code languages, tools, and services used | python, CUDA |
| Compilation requirements, operating environments & dependencies | Linux, OSX & Python Scientific Computing Stack |
| If available Link to developer documentation/manual | For example: https://usgs-astrogeology.github.io/autocnet/ |
| Support email for questions | jlaura@usgs.gov |

(Source: https://doi.org/10.1016/j.softx.2018.02.001)

**4.5 What reasons are given for referencing Jupyter notebooks in publications?**

In identifying the location of links and citations in each of the 392 papers, we documented any reasons given for using a Jupyter notebook. Rarely were these more than mentions in passing, if any reason given it all, so a definitive coding scheme was infeasible. Among the variety of tasks or reasons the authors did mention were data analysis, reproducibility, tutorials, and integration with one or more products developed by the author. Many papers cited Jupyter notebooks as tutorials for products developed by the authors. For instance, a software article on the Colossus Python package uses Jupyter notebooks as "tutorials that explain how each module" of the package can be used [22]. They provide both static html versions of these notebooks and editable notebooks.

Several others stated that they used Jupyter notebooks to foster reproducibility of their own research results, often using Peng's definition of computational reproducibility as the availability of data and code used in the analysis [23]. For example, an astronomy article examining masses of binary stars links to Jupyter notebooks on GitHub to promote "open and reproducible research" [24]. Their GitHub repository contains eight Jupyter notebooks, .csv files of associated data, Python scripts, and the paper itself. These notebooks each show an analysis performed to obtain the paper's results with comments that explain the functionality of the code and software. These notebooks are only "partially reusable" in that the code in the notebooks is available with the caveat that some of the code links to data on the author's hard drive, with the implication that the users would supply their own data. In this example, the researchers are using Jupyter notebooks to provide code and software enumerating the procedures and processes to refine the data.

## 5 DISCUSSION

Our study of data and software citations in the astronomy literature to Jupyter notebooks, from their inception in 2014 through 2018, reveals rapid adoption of these tools. However, our findings also demonstrate that mentions of Jupyter notebooks rarely provide the credit, attribution, and discovery of notebooks intended by data and software citation principles. Fewer than half of the mentions lead to findable and accessible notebooks, and of these, only about one fourth can be considered citations. Thus, about 12% of 897 Jupyter mentions in five years of astronomy literature include citations to notebooks. Most (67%) of the notebooks found are located in GitHub repositories.

The majority of references to Jupyter notebooks consist only of links without associated metadata. Of these, links to project websites are most subject to link rot, thus exhibiting the least fixity. Citations sometimes occur in forms analogous to bibliographic citations, as recommended in the data and software citation principles, but they also appear as tables of metadata. Links and citations to Jupyter notebooks are scattered throughout the text of papers, in footnotes, in appendices, and in supplemental material. In sum, Jupyter notebooks may serve their intended purposes of aggregating research objects and improving reproducibility of research, but appear to be doing little to tame the morass of uncited data and software.

Jupyter notebooks, at least in the astronomy case study for which we have evidence, are not addressing the fixity problems of citation. Notebooks stored in GitHub or similar personal repositories do not appear to be stabilized as fixed objects that can be retrieved in the same form at a later time. Rather, GitHub repositories, by design, are flexible work areas that can be reorganized frequently. In only a few cases did links or citations to Jupyter notebooks point to data or software repositories that curate research objects for long-term fixity and findability.

Similarly, Jupyter notebooks do not appear address the granularity problems of data and software citation because most are referenced with links alone, failing to provide enough metadata to describe the boundaries of a research object. The notebook itself is a research object, and one that provides context and relationships between the objects it contains. In considering matters of fixity and granularity, it is important to consider the location of Jupyter notebooks at the top of Hinson's software stack. The stability of any notebook depends on the stability of the software layers on which it rests. The more objects and relationships within a notebook, the more unstable it may become over time. "Software collapse" of Jupyter notebooks looms large.

## 6 CONCLUSION AND RECOMMENDATIONS

Jupyter notebooks are popular in astronomy and other scientific fields because they serve multiple functions in conducting research such as documenting data analysis, aggregating disparate but related objects, and providing tutorials on software and other tools. As presently deployed, these notebooks are more valuable for the research process than for facilitating access to research products. We found little evidence that Jupyter notebooks provided credit or attribution for datasets and software, or that they facilitated access to individual research objects. The answer to our framing question, of whether Jupyter notebooks are being used to discover, access, and give credit to the objects they contain, or whether today's notebooks "black box" those objects, is the latter.

If Jupyter notebooks are not the answer to the credit, attribution, discovery, fixity, and granularity problems in endemic in providing access to research data and software, what is? Our analysis leads to several recommendations that address these problems, if not to full answers.

Researchers, as authors and as users of tools such as Jupyter notebooks, can give credit where due by citing datasets and software tools individually, in addition to the notebooks in which they may reside. Any objects referenced should be described with adequate metadata, following the data and software citation principles, and should be deposited in a fixed form at a stable location. Similarly, when referencing GitHub repositories, those repositories and notebooks should be fixed and stabilized to serve the functions of credit, attribution, discovery, and reproducibility.

Publishers can play important roles in improving data and software citation, and thus giving credit and facilitating access. One means is to allow longer citation lists that can include references to data and software. Journals, that limit references to 20 or fewer are creating disincentives to make these additional citations. Another tool available to publishers is the choice of citation styles. Styles that maximize metadata and include persistent and unique identifiers such as DOIs facilitate credit, attribution, and discovery of all research objects, including data and software. A surprising array of citation styles are incomplete, removing unique and persistent identifiers such as DOIs, and introducing ambiguity by abbreviating journal names. Funding agencies can improve data and software citation by encouraging researchers and publishers to follow these recommendations.

Jupyter notebooks are useful tools as living research objects. Notebooks that are to serve citation functions, however, must be frozen as fixed objects. If cited, they should not be moved, as link rot sets in quickly. Citations within Jupyter notebooks should follow recommended best practices for data and software and utilize existing extensions that store references in a notebook's metadata. However, Jupyter notebooks are no more stable in the long term than any other information technology. Their stability can be improved by deploying relevant extensions that facilitate notebooks' use as access points to discover other research objects. Best practice overall is to provide full citations, not simply links, to Jupyter notebooks and to each of the datasets and software objects they contain, which will give credit where credit is due.

## ACKNOWLEDGMENT

This research is funded by the Alfred P. Sloan Foundation Award #2015-14001: If Data Sharing is the Answer, What is the Question? We thank Michael Scroggins, Peter Darch, and Cheryl Thompson of the UCLA Center for Knowledge Infrastructure for their comments on drafts. We also gratefully acknowledge use of NASA's Astrophysics Data System Bibliographic Services.

**Morgan F. Wofford** completed her MLIS degree in 2018 in the UCLA Information Studies Department. She earned a Bachelor of Science Degree in Conservation Biology from the University of California, Davis in 2014. She is currently the UCLA Center for Knowledge Infrastructures' Data Scientist and Project Manager. She is a member of the Association for Information Science and Technology. She is especially interested in data curation, open science, and information policy.

**Bernadette M. Boscoe** completed her PhD in 2019 in the UCLA Information Studies Department. Dr. Boscoe is currently a member of the Center for Knowledge Infrastructures at UCLA. She has accepted a Postdoctoral Fellowship at the School of Information at the University of Washington. Her research concerns keeping scientific data alive over decades. She brings her experience in computer science and mathematics to studies of data and software relationships.

**Christine L. Borgman** holds the PhD in Communication from Stanford University, MLS from Pittsburgh, and a BA in Mathematics from Michigan State. A Distinguished Research Professor and Director of the Center for Knowledge Infrastructures at UCLA, she is the author of more than 250 publications in information studies, computer science, communication, and law, including three books from MIT Press: Big Data, Little Data, No Data: Scholarship in the Networked World (2015), Scholarship in the Digital Age: Information, Infrastructure, and the Internet (2007); and From Gutenberg to the Global Information Infrastructure: Access to Information in a Networked World (2000). She is a Fellow of the American Association for the Advancement of Science and of the Association for Computing Machinery; she co-chaired the CODATA-ICSTI Task Group on Data Citation and Attribution.

**Irene V. Pasquetto** completed her PhD in 2018 in the UCLA Information Studies Department. She earned a Bachelors Degree in Communication Science and a Masters Degree in Journalism and Publishing Industry both from Verona University (Italy). During her time at UCLA, she worked at the Center for Knowledge Infrastructures. Dr. Pasquetto is now a Postdoctoral Fellow at the Shorenstein Center of the Kennedy School of Government, Harvard University. She has accepted a tenure-track assistant professor position at the University of Michigan School of Information beginning in August 2020. Irene's research interests included open knowledge, open data and open science, especially their policy and economic implications.

**Milena S. Golshan** holds MLIS degrees from UCLA and Sofia University St. Kliment Ohridski in Bulgaria. She was the UCLA Center for Knowledge Infrastructures' Data Scientist and Project Manager. Ms. Golshan is now a Metadata Specialist in Collections Management at the Los Angeles County Museum of Art. Milena's experience includes academic research support, qualitative data analysis, content management, and information architecture. She is especially interested in open data, digital archiving, data analysis, management, and curation.